\documentstyle[preprint,aps]{revtex}
\begin{document}

\title {THE CONDUCTANCE IN THE ONE DIMENSIONAL SPIN POLARIZED  GAS} 
\author{D.Schmeltzer}
\address{Department of  Physics City  College  of New-York, 
N.Y.10031, U.S.A }
\author{E. Kogan, R. Berkovits and M. Kaveh}
\address{Minerva Center, Jack and Pearl Resnick Institute 
of Advanced Technology, Department  of   Physics,  Bar Ilan University
Ramat-Gan,  ISRAEL}
\date{\today}
\maketitle
\begin{abstract}
 We present a theoretical analysis of the recent experimental  
results of Thomas {\it et al} on transport properties of spin polarized  
quantum wires. We suggest an explanation of the observed deviations of  
the conductance from the universal value $G=2e^2/h$ per channel in the  
wire. We argue that the new quasi plateau observed for the conductance  
at the value $G=1.4e^2/h$  is a result of the proximity between the spin  
polarized phase and the metallic one. The  enhancement of the  
conductance from the value $G=e^2/h$  to $G=1.4e^2/h$  is due to the  
hybridization of the electronic state at   $K_F^{\rm \downarrow}\approx 0$  
with the chiral states at $K_F^{\rm \uparrow}$    
and $-K_F^{\rm \uparrow}$.
\end{abstract}    
\pacs{PACS numbers: } 

Recent experiments on quantum wires show discrepancies  
between the ballistic conductance measured by different groups
\cite{tarucha,yacoby,thomas}. 
According to the Tomonaga-Luttinger theory  the ballistic 
conductance is normalized by the electron-electron interaction  
$G=K_c 2e^2/h$, where $K_c<1$  for  $\omega<v_c/L$ and $K_c =1$ for  
$\omega>v_c/L$ ($\omega$ is the frequency, $v_c$
is the charge density velocity  and $L$ is length of the wire). Contrary to
these results, Shimizu \cite{shimizu} has argued that within the Landauer
\cite{landauer} theory the
conductance remains unrenormalized. This  result was understood by
Finkelstein \cite{finkelstein} and Kavabata \cite{kavabata}  as a
renormalization of both the current and the driven voltage. It is  important 
to mention  that the  experimental and theoretical work has been restricted 
to cases where complete charge-  spin  separation exists. The Cavendish  group
experiment \cite{thomas} is new in the sense that it raises a new question, 
namely 
ballistic transport in the presence of  weak coupling between charge and 
spin excitations. We offer an explanation  to this experiment and analyze for
the first  time the conductance  in  the presence  of spin charge coupling.  
	
The main features of the experiment are:

a)At  low temperatures as the gate voltage is varied, the
conductance shows a number of plateaus separated by the steps  $G=2e^2/h$.
In addition to those plateaus, a new quasi plateau with the value  of the
conductance  $G=0.7\times 2e^2/h$ is observed. 

b)With the increase of the external magnetic field, the new quasi 
plateau shifts to lower values approaching $G=e^2/h$. This corresponds to
the conductance of a single polarized channel.
 
c)As the temperature increases from 0.07 K to 1.5 K, the quasi
plateau value of the conductance  decreases. 

d)The  lack of inversion symmetry and the presence of interface
electric field, induce zero field spin splitting in GaAs/AlGaAs
heterostuctures. The splitting leads to the lift of the spin degeneracy and
creates spin polarization sub-bands in a zero magnetic field. 

Based on the experimental  facts, we suggest the following
explanation: 
	 
The two dimensional  GaAs electron  gas is polarized at  T=0. Due
to the fact that  the GaAs electron mass is small  $m=0.08m_e$ and $g=0.5$,
it
follows that in the presence of a magnetic field the Zeeman term is
negligible with respect to the orbital  motion. The lack of inversion
symmetry allows  the spin orbit coupling  which lifts the spin
degeneracy. As a result, the spin split is much larger than the Zeeman
term would give. The presence of the orbital motion induces the coupling
between the one dimensional modes when the electron motion is
squeezed by a transversal electrostatic potential. Effectively we describe
the spin polarization in the quantum wire by a field  $h$, where 
$h=g\mu B+R\left(N^{\uparrow} - N^{\downarrow}\right)
/\left(N^{\uparrow} + N^{\downarrow}\right)$. The second part of $h$ 
includes the two dimensional many
body effect which induces spin polarization at $T=0$, so $R$ is, in fact, an 
effective exchange constant. 
	At low gate voltages only one mode is propagating. Due to the
spin polarization only one component spin  (spin  up) is propagating,
leading to the conductance of one channel, $G=e^2/h$. This  indeed is seen
in the experiment at large magnetic fields. When  the magnetic field
decreases, the polarization  gap is reduced. Due to electron-electron
interaction the propagating mode (spin up ) is hybridized with the spin
down mode. This enhances the conductance. We suggest that the quasi
plateau at  $G=1.4\times e^2/h$ occurs when the gate voltage is such that the
propagating mode  with the Fermi surface  at $\pm K_F^{\uparrow} \neq 0$ 
is degenerated with 
the  state $ K_F^{\downarrow} = 0$. Increasing further the gate voltage, 
we obtain  two
propagating modes with four Fermi surfaces at  $\pm K_F^{\uparrow} \neq 0$
and   $\pm K_F^{\downarrow} \neq 0$.      
In this case the system behaves like a charged Luttinger liquid weakly
coupled to the spin liquid. As a result we find that the conductance is
$G=K_{\rm eff}\times 2e^2/h$ with $K_{\rm eff}\sim K_c$ ($K_{\rm eff}$ is
renormalized by the spin liquid). If the arguments presented in ref.
\cite{shimizu,finkelstein,kavabata} hold for our case, we would expect to
find  $G=2e^2/h$. 

Formally we show this within the Hubbard model in the presence
of a fixed magnetic field h and a tunable chemical potential $E_F$
(by the gate voltage). 

In the presence of the magnetic field h we identify the following
cases: 

\noindent
a) The two (spin up and spin down ) one dimensional modes are
propagating. The value of the  Fermi Surface points are given by the
solution $\epsilon\left(\pm K_F^{\rm \uparrow}\right) -E_F  - h/2 =0$
and $\epsilon\left(\pm K_F^{\rm \downarrow}\right) -E_F  + h/2 =0$ with 
$K_F^{\rm \uparrow}\neq 0$ and $K_F^{\rm \downarrow}\neq 0$.

\noindent 
b)The polarized case, 
$\epsilon\left(\pm K_F^{\rm \uparrow}\right) -E_F  -h/2 =0$ 
and  no Fermi surface for the spin down band. 

\noindent
c)The  phase transition case, characterized  by the degeneracy of the three 
Fermi surfaces: $K_F^{\rm \downarrow}=0$, $+K_F^{\rm \uparrow}\neq 0$ and 
$-K_F^{\rm \uparrow}\neq 0$. 

Let us consider the cases in details.

\subsection{Two propagating modes -- Isotropic case}

We consider the one dimensional Hubbard model in the presence
of a weak magnetic field h. Within the one dimensional Bosonization we 
find  a weakly coupled charge and spin liquid. 
The charge  is described
by the bosonic field $\theta_c$, canonical momentum $P_c$, 
charge density velocity $v_c$
and charge stiffness $K_c  <1$. The spin liquid is described by the bosonic
field $\theta_s$, canonical momentum $P_s$, spin density velocity $v_s$
and spin stiffness $K_s >1$. 
	\begin{eqnarray}
	\label{hamiltonian}
	H&=&\int dx\left\{\frac{v_c}{2} \left[K_c P^2_c + 
	\frac{\left(\partial_x \theta_c\right)^2}{K_c}\right]+
	\frac{v_s}{2} \left[K_s P^2_s + 
	\frac{\left(\partial_x \theta_s\right)^2}{K_s}\right] \right.  
	\nonumber \\	
	&& \left. +h\left[P_c P_s+
	\partial_x \theta_c \cdot\partial_x \theta_s\right]+
	\sqrt{\frac{2}{\pi}}V^{\rm ext}(x,t)\partial_x \theta_c\right\}
	\end{eqnarray}
where $v_c =v_F(1+g/\pi v_F )^{1/2}$,  $v_s =v_F(1-g/\pi v_F )^{1/2}$, 
$K_c =(1+g/\pi v_F )^{-1/2}$, $K_s =(1-g/\pi v_F )^{-1/2}$, 
$v^{\uparrow}_F=v_F+h$, $v^{\downarrow}_F=v_F-h$, $v_F =K_F/m$. 
The potential $V^{\rm ext}(x,t)$ is the external scalar potential
introduced to probe the system. The effect of the weak magnetic field
gives rise to a spin charge coupling. The conductance for this case will
be $G=K_{\rm eff}\times 2e^2/h$. If the arguments given in
ref.\cite{shimizu,finkelstein,kavabata} will hold here, we expect
to find $G=2e^2/h$, in agreement with the plateau observed in the
experiment. 

\subsection{Anisotropic case} 

When the electrons are polarized, only one mode is
propagating, $C_{\uparrow}(x) =C(x)$. The second mode  $C_{\downarrow}(x)=
\psi(x)$ is characterized by the spin gap $D>0$, $D=h/2-E_F$. Increasing the
gate voltage  the spin gap vanishes inducing an enhancement of the
conductance. When $D<0$ we have the anisotropic case with 
$K_F^{\uparrow}\gg K_F^{\downarrow}$. The Hamiltonian is: 
	\begin{eqnarray}
	\label{hamiltonianb}
	H&=& \int dx\left\{C^{\dagger}(x)\left(-\frac{\partial_x^2}{2m}
	-E_F-\frac{h}{2}\right)C(x)+
	\psi^{\dagger}(x)\left(-\frac{\partial_x^2}{2m}-E_F+\frac{h}{2}\right)
	\psi(x) \right.\nonumber \\
	&&\left. +g:c^{\dagger}(x)c(x):\psi^{\dagger}(x)\psi(x)
	+V^{\rm ext}(x,t)
	\left(:C^{\dagger}(x)C(x):+ \psi^{\dagger}(x)\psi(x)\right)
	\right\},
	\end{eqnarray}
where  $g$ is the Hubbard interaction and $V^{\rm ext}(x,t)$ is the external
potential. 

We bosonize the metallic electrons: 
	\begin{equation}
	\label{boson}
	C(x)=\frac{1}{\sqrt{2\pi a}}\left[e^{iK_F^{\uparrow}x}:
	e^{i\sqrt{4\pi}\theta_{+}}:+e^{-iK_F^{\uparrow}x}
	:e^{-i\sqrt{4\pi}\theta_{-}}:\right]
	\end{equation}
where $\theta_{+}+\theta_{-}=\theta$, $\theta_{-}+\theta_{+}=\phi$, 
$v_F^{\uparrow} =v_F +h$, $v_F =K_F/m$.  The bosonic density 
$:C^{\dagger}(x)C(x):=1/\sqrt{\pi}\partial_{x}\theta+\frac{1}{\pi a}
\cos\left(2K_F^{\uparrow}x+ \sqrt{4\pi}\theta(x)\right)$. The Euclidean
action  $S$ for the Hamiltonian  (\ref{hamiltonian}) is 
 	\begin{eqnarray}
 	\label{action}
	S&=&\int_0^{\beta}d\tau\int
	dx\left\{\frac{1}{2}\left[\frac{1}{v_F^{\uparrow}}
	\left(\partial_{\tau} \theta\right)^2 +v_F^{\uparrow}
	\left(\partial_{x} \theta\right)^2\right] -
	\frac{\partial_x\theta}{\sqrt{\pi}}\left[g\psi^{\dagger}(x)\psi(x)
	-iV^{\rm ext}(x,\tau)\right]\right. \nonumber \\
	&&\left.
	+ \psi^{\dagger}\left(x,\tau\right)
	\left[\left(\partial_{\tau}-E_F+\frac{h}{2} + iV^{\rm
	ext}\left(x,\tau\right)\right)-
	\frac{\partial^2_x}{2m}\right]\psi(x,\tau) \right\}
	\end{eqnarray}
In the Eq. (\ref{action}) we have ignored the oscillatory term  
$\cos\left(2K^{\dagger}_F x + \sqrt{4\pi}\theta(x)\right)$. We
compute  the generating function  $W\left(V (x,t)\right)$
which will be used to compute the current. 
 	\begin{equation}
 	\label{partition}
	Z=\int D \psi^{\dagger}D\psi D\theta \exp\{-S\}=
	\int D \psi^{\dagger}D\psi D a_0 \exp\{-\tilde{S}\}=
	\exp\left\{-W\left(V^{\rm ext}\right)\right\}
	\end{equation}
with
 	\begin{equation}
 	\label{action2}
	\tilde{S}=\tilde{S}\left(V^{\rm ext}\right)+
	\tilde{S}\left(\psi^{\dagger},\psi\right)+\tilde{S}\left(a_0\right).
 	\end{equation}
 	
The integration of the bosonic variables $\theta$ induces an effective 
interaction $\tilde{U}(q,\omega_n)$ between the $\psi(x)=C_{\downarrow}(x)$ 
fermions. This induced two body interaction is replaced by the action 
$\tilde{S}\left(a_0\right)$  with the auxiliary scalar field $a_0$.
The action in the Eq. (\ref{action2}) 
was obtained after the bosonic field $\theta$ was
integrated . Using the Matsubara frequencies  $\omega_n =2\pi T n$ 
for the bosons and $\nu_n =2\pi T(n+1/2)$ for the fermions we obtain: 
 	\begin{equation}
 	\label{action3}
	\tilde{S}\left(V^{\rm ext}\right)=\sum_{\omega_n}\sum_{q}\frac{1}{2\pi
	v_F^{\uparrow}}V^{\rm ext}\left(q,\omega_n\right)
	\frac{\left(v^{\uparrow}_Fq\right)^2}
	{\left(v^{\uparrow}_Fq\right)^2+\omega_n^2}
	V^{\rm ext}\left(-q,-\omega_n\right)
 	\end{equation}
	\begin{eqnarray}
 	\label{action4}
	\tilde{S}\left(\psi^{\dagger},\psi\right)=\sum_{\nu_m}\sum_p
	\left\{-\psi^{\dagger}\left(p,\nu_n\right)
	\left[-\nu_n-\frac{p^2}{2m}+\left(E_F+\frac{g^2}{2\pi
	v_F^{\uparrow}}-\frac{h}{2}\right)\right]\psi\left(p,\nu_n\right)
	\right.\nonumber \\
	\left.+i\sum_{\omega_n}\sum_{q}\psi^{\dagger}
	\left(p+q,\nu_n+\omega_n\right)\left[V^{\rm 		   
	ext}\left(q,\omega_n\right)
	\left(1-\frac{g}{\pi v^{\uparrow}_F}
	\frac{\left(v^{\uparrow}_Fq\right)^2}
	{\left(v^{\uparrow}_Fq\right)^2+\omega_n^2}\right)
	+a_0\left(q,\omega_n\right)\right]\psi\left(p,\nu_n\right)\right\}
 	\end{eqnarray}
	\begin{equation}
 	\label{action5}
	\tilde{S}\left(a_0\right)=\sum_{\omega_n}\sum_{q}\frac{1}{2}
	a_0\left(q,\omega_n\right)\tilde{U}^{-1}\left(q,\omega_n\right)
	a_0\left(-q,-\omega_n\right)
	\end{equation}
	\begin{equation}
	\label{potential}
	\tilde{U}(q,\omega_n)= \frac{g^2}{\pi v^{\uparrow}_F}
	\left(\frac{\omega_n^2}
	{\omega_n^2+\left(v^{\uparrow}_F q\right)^2}\right)
	\end{equation}	

The interaction in Eq. (\ref{action5}) is induced by the integration of the 
bosonic field $\theta$ (the upper band electrons).
Due to the Hubbard interaction  the  single particle energy of the
localized  band is shifted down, $\epsilon(q)\rightarrow
\epsilon(q)-g^2/2\pi v_F^{\uparrow}$, $\epsilon(q)=q^2/2m$. The
new gap function will be: 
	\begin{equation}
	\label{gap} 
	\Delta=h/2-E_F -g^2/2\pi v^{\uparrow}_F. 
	\end{equation} 

As a result the polarized state will exist in strong magnetic fields  such
that $D>0$, $\Delta>0$.  In the range $D>0$ and $\Delta<0$  we obtain the
hybridized  state. 

\subsection{The hybridized  state}

The integration of the fermion field induces an effective action for this
field. Keeping only second order terms in the auxiliary field we
obtain the generating  function $W\left(V (x,\tau)\right)$. 
	\begin{equation}
	\label{partition2}
	Z=\exp\left(-W\left(V^{\rm ext}(x,\tau)\right)\right)                 
    	\end{equation}		                                              				       			 
 	\begin{eqnarray}
 	\label{generator}
	&&W\left(V^{\rm ext}(x,\tau)\right)= \sum_{\omega_n}\sum_{q}
	\frac{1}{2\pi}V^{\rm ext}\left(q,\omega_n\right) \nonumber \\ 
	&&\left[\frac{v^{\uparrow}_F q^2}
	{\left(v^{\uparrow}_Fq\right)^2+\omega_n^2}+
	\left(1-\frac{g}{\pi v^{\uparrow}_F} \cdot 
	\frac{\left(v^{\uparrow}_Fq\right)^2}
	{\left(v^{\uparrow}_Fq\right)^2+\omega_n^2}\right)^2
	\Pi\left(q,\omega_n\right)\right]
	V^{\rm ext}\left(-q,-\omega_n\right).
 	\end{eqnarray}
 
In  Eq. (\ref{generator}) $\Pi(q, \omega)$ is the non-interacting polarization 
diagram for the spin down polarized electrons. We investigate Eq. 
(\ref{generator}) at $T=0$ and $T\neq 0$. 
	\begin{equation}
	\label{polarization}	 
	\Pi\left(q,\omega_n\right)= \int\frac{dP}{2\pi}
	\frac{f\left(\hat{\epsilon}(p)\right)-
	f\left(\hat{\epsilon}(p+q)\right)}
	{i\omega_n-\hat{\epsilon}(p+q)+\hat{\epsilon}(q)}
	\end{equation}
Where $\tilde{\epsilon}(p)=\epsilon(p) + \Delta$, $\epsilon(p)=p^2/2m$ 
and $f(\epsilon + \Delta )= \left(\exp(\epsilon + \Delta)/T +1\right)^{-1}$ 
is the Fermi-Dirac function.  At $T\neq 0$ and large magnetic fields such that 
$\Delta >0$ the polarization diagram obeys
	\begin{equation}
	\label{v}
	v^{\uparrow}_F \pi(q,\omega_n;\; \Delta > 0 )=
	\left(v^{\uparrow}_F/v^{\uparrow}_0\right)
	\exp\left(-\Delta/T\right),\; v^{\downarrow}_0 = \sqrt{T/m} 
	\end{equation}  
For $D >0$ and $\Delta <0$	we have                           
    	\begin{equation}
 	\label{v2}
	v^{\uparrow}_F \pi(q,\omega_n;\; \Delta < 0 )=
	\left(v^{\uparrow}_F/v^{\downarrow}_0\right), \;\;
	v^{\downarrow}_0= \sqrt{2\left(E_F +g^2/2\pi v_F -h/2\right)/m}    
    \end{equation}
 
In order to compute the current from  the generating  function
$W\left(V(x,\tau)\right)$, we perform the analytic continuation 
$\omega_n \rightarrow i\omega -x$   in Eq. (\ref{generator}). 
The current is given by: 
	\begin{equation}
	\label{current}
	I(q,\omega)=\frac{\omega}{q}\frac{\partial} 
	{\partial V^{\rm ext}} W\left(V^{\rm ext}\right)
	\end{equation}
 
The  conductance is obtained from Eq. (\ref{current})  by taking the limit 
$\omega \rightarrow 0$, $q \rightarrow 0$.  The conductance  $G(T)$
in the units  $e=\hbar=1$, $1/2\pi=e^2/2\pi\hbar =e^2/h$ is
	\begin{equation}
	\label{conductance}
	G(T)=e^2/h\left(1 +\left(1-g/\pi v_F^{\uparrow}\right)^2  
	\left(v_F^{\uparrow}/v_0^{\downarrow}\right)
	F(T)\right)
        \end{equation}
Where $F(T)=exp(-\Delta/T)$   for   $\Delta>0$  and  $F(T)=1$    for
$\Delta<0$   and $T\rightarrow 0$.  From the Eq. (\ref{conductance}) we observe 
that for  large
magnetic fields  $\Delta\gg T$, $F(T)\rightarrow  0$.As a result $G(T)=e /h$ 
in agreement with the experiment. Decreasing the magnetic field, one finds
that $G(T)$ increases:    
	\begin{equation}
	\label{conductance2}
	G(T)=e^2/h\left(1+\left(1-g/\pi v_F^{\uparrow}\right)^2
	\left(v_F^{\uparrow}/\sqrt{T/m}\right) \exp(-\Delta/T)\right). 
	\end{equation}

In the last part we compute the conductivity $\sigma\left(\omega,q\right)$ at 
$T=0$ as a function of the Hubbard interaction $g$ and the gap function $\Delta$ 
given in Eq. (\ref{gap}). We replace in Eq. (\ref{generator},\ref{polarization}) 
$\omega_n$ for $i\omega -x$ and 
find for the conductivity
$\sigma\left(\omega,q\right)$
	\begin{eqnarray}
	\label{sigma}
	\sigma(\omega,q)=\frac{\pi e^2}{h}\left\{v_F^{\uparrow}
	\left(\delta\left(\omega-v_F^{\uparrow}q\right)+
	\delta\left(\omega+v_F^{\uparrow}q\right)\right) \right.\nonumber\\
	\left.
	+\mu(-\Delta)\left[v_0^{\downarrow}
	\left(\delta\left(\omega-v_0^{\downarrow}q\right)+
	\delta\left(\omega+v_0^{\downarrow}q\right)\right)
	+\frac{2g}{\pi}\left(\frac{v_0^{\downarrow}v_F^{\uparrow}}
	{(v_F^{\uparrow})^2-(v_0^{\downarrow})^2}\right)\right.
	\right.\nonumber\\
	\left.\left.
	\left(\delta\left(\omega-v_0^{\downarrow}q\right)+
	\delta\left(\omega+v_0^{\downarrow}q\right)
	-\delta\left(\omega-v_F^{\uparrow}q\right)-
	\delta\left(\omega+v_F^{\uparrow}q\right)\right)\right]\right\}
	\end{eqnarray}
In Eq. (\ref{sigma}) we have $\mu(-\Delta)=0$ for $\Delta > 0$ and 
$\mu(-\Delta)=1$	for $\Delta \leq 0$, 
$v_0^{\downarrow}=\mu(-\Delta)\sqrt{\frac{2}{m}(-\Delta)}$ and 
$v_F^{\uparrow}=v_{F}+h$.

Using Eq. (\ref{sigma}) we find that for an infinite system at $\omega=0$ the 
conductance $G$ is given by
	\begin{equation}
	G=\left\{ 
	\begin{array}{cc}
	\frac{e^2}{h}, & \Delta>0\\
	\frac{2e^2}{h}, & \Delta\leq0
	\end{array}
	\right.
	\label{con}
	\end{equation}
For finite frequencies and	finite samples we can have $\frac{e^2}{h}\leq G\leq 
\frac{2e^2}{h}$. We suggest that the quasi-plateau observed at $G=1.4 e^2/h$ can 
be explained by assuming a strong anisotropy for the velocities 
$v_0^{\downarrow}\ll v_F^{\uparrow}$. This is achieved for $D=0$, 
$\Delta = -\frac{g^2}{2\pi v_F^{\uparrow}}$.
For a finite sample of length $L$ the conductivity at a finite frequency 
$\omega$ is determined by the lowest mode $v_0^{\downarrow}\frac{2\pi}{L}$, we 
find:	
	\begin{equation}
	\label{conductance3}
	G=e^2/h\left[1+\frac{2g}{\pi v_F^{\uparrow}} 
	\left(1-\left(\frac{v_0^{\downarrow}}{v_F^{\uparrow}}
	\right)^2\right)^{-1}\right]
	\simeq \frac{e^2}{h}K; \;\;\; 
	v_F^{\uparrow}\frac{2\pi}{L}\gg\omega
	\geq v_0^{\downarrow}\frac{2\pi}{L}
	\end{equation}
where 
$K\simeq\left[1-\frac{4g}{\pi v_F^{\uparrow}}\right]^{-1/2} >1$ 
is the interaction parameter for an attractive interaction generated by the 
electrons with opposite polarization $\left(K_F^{\uparrow}\simeq 0\right)$. 
This result follows, in fact, from the effective attractive interaction which
originated from the repulsive interaction of the electrons with opposite spins. 
Eq. (\ref{conductance3}) suggests a possible qualitative explanation to the 
experimental value of  $G=1.4 e^2/h$. 

To conclude, we identify the quasi-plateau
observed by the Cavendish group  with the strong hybridization between
the "up" and "down" electrons.

We are particularly grateful to  M. Pepper for helpful discussions.

\end{document}